\newcommand{\beq}{\begin{equation}}
\newcommand{\eeq}{\end{equation}}
\newcommand{\bea}{\begin{eqnarray}}
\newcommand{\eea}{\end{eqnarray}}

\newcommand{\gsim}{\lower.7ex\hbox{$\;\stackrel{\textstyle>}{\sim}\;$}}
\newcommand{\lsim}{\lower.7ex\hbox{$\;\stackrel{\textstyle<}{\sim}\;$}}
\newcommand{\nnmb}{\nonumber}

\documentclass[aps,prev,twocolumn,preprintnumbers,%
               floatfix,nofootinbib]{revtex4}
\usepackage{graphicx}
\usepackage{mathrsfs}
\usepackage{amssymb}
\usepackage{verbatim}

\def\stacksymbols #1#2#3#4{\def\theguybelow{#2}
    \def\vp{\lower#3pt}
    \def\sp{\baselineskip0pt\lineskip#4pt}
    \mathrel{\mathpalette\intermediary#1}}

\def\intermediary#1#2{\vp\vbox{\sp
     \everycr={}\tabskip0pt
     \halign{$\mathsurround0pt#1\hfil##\hfil$\crcr#2\crcr
              \theguybelow\crcr}}}

\def\comment#1{}
\def\to{\rightarrow}

\def\u1x{U(1)_X}
\newcommand{\nc}{\newcommand}
\nc{\LL}{L}
\nc{\vv}{\tilde{v}}
\nc{\ccdot}{\!\cdot\!}
\nc{\gsm}{G_{SM}}
\nc{\vfive}{\mathbf{5}\oplus\mathbf{\overline{5}}}
\nc{\vten}{\mathbf{10}\oplus\mathbf{\overline{10}}}

\begin{document}

%
%

\preprint{MCTP-05-101}

\title{The tension between gauge coupling unification, the Higgs boson mass,\\
and a gauge-breaking origin of the supersymmetric $\mu$-term}

\author{David E. Morrissey}
\email{dmorri@umich.edu}
\author{James D. Wells} 
\email{jwells@umich.edu}
\vspace{0.2cm}
\affiliation{
Michigan Center for Theoretical Physics (MCTP) \\
Department of Physics, University of Michigan, Ann Arbor, MI 48109}
\date{\today}

\begin{abstract}

We investigate the possibility of generating the $\mu$-term in the
MSSM by the condensation of a field that is a singlet under the SM 
gauge group but charged under an additional family-independent 
$\u1x$ gauge symmetry.  We attempt to do so while preserving the  
gauge coupling unification of the MSSM.  For this, we find that 
SM non-singlet exotics must be present in the spectrum. We also prove 
that the pure $U(1)_X$ anomalies can always be solved with rationally 
charged fields, but that a large number of SM singlets are often required.  
For $U(1)_X$ charges that are consistent with an 
embedding of the MSSM in $SU(5)$ or $SO(10)$, we show that the $U(1)_X$ 
charges of the MSSM states can always be expressed as a linear combination 
of abelian subgroups of $E_6$.  However, the SM exotics do not appear 
to have a straightforward embedding into GUT multiplets.  We conclude from 
this study that if this approach to the $\mu$-term is correct, 
as experiment can probe, it will necessarily complicate the standard
picture of supersymmetric grand unification.

\end{abstract}

\maketitle


\maketitle


\setcounter{equation}{0}


\section{Introduction}

  The minimal supersymmetric standard model (MSSM) has two 
problems associated with its Higgs boson sector.  The first has
to do with the $\mu$-term in the superpotential, 
$W_{MSSM} \supset \mu\,H_d\ccdot H_u,$ which must be of order the 
electroweak scale, $\mu \sim v = 174$~GeV, for electroweak symmetry
breaking to occur at the correct scale in a natural way.  
This is also true of the scale of supersymmetry~(SUSY) breaking.
However, from the low-energy point-of-view, $\mu$ and the SUSY breaking
scale are independent of each other.  Even though both scales are
stable under quantum corrections because of supersymmetry,
it is curious that their numerical values should be so close to each other,
or why $\mu$ is so much smaller than the seemingly more fundamental
scales $M_{\rm{GUT}}$ or $M_{\rm{Pl}}$~\cite{Giudice:1988yz}.  
   
  The second problem is related to the first.  At tree-level, the mass
of the lightest CP-even MSSM Higgs boson is bounded by the mass
of the $Z^0$ gauge boson, $m_h^2 \leq M_Z^2\cos^2\beta$.
This is considerably less than the current experimental bound from
LEP~\cite{Barate:2003sz}; $m_h \gtrsim 114~\mbox{GeV},$
for a SM-like $h^0$.  The experimental bound on the Higgs mass does
not rule out the MSSM because this quantity can receive large radiative
corrections, especially from a heavy scalar top, provided the SUSY
breaking scale is larger than about 
$M_{SUSY} \gtrsim 1$~TeV~\cite{Carena:2002es}.
But, since the electroweak scale is related (schematically) 
to $\mu$ and $M_{SUSY}$ by the relation 
$v^2 \sim \left|\mu^2 - M^2_{SUSY}\right|$,
such a large SUSY breaking scale requires a fine-tuning of 
$\mu$ at the percent level.  Again, this fine-tuning
is not disastrous, but it is not terribly appealing in a model 
that was motivated by naturalness in the first place.  

  One way to avoid both of these problems is to add a singlet chiral
superfield to the MSSM.  With a singlet, the $\mu$-term in the 
superpotential can be replaced by $W \supset \lambda\,S\,H_d\!\cdot\!H_u$,
giving $\mu_{eff} = \lambda\,\left<S\right>$.  Since the vacuum expectation
value~(VEV) of the singlet scalar is largely determined by the associated
soft SUSY breaking terms, this replacement relates $\mu_{eff}$ to
$M_{SUSY}$, and thus explains the coincidence of these scales.
The singlet term also helps to remove the fine-tuning required by 
the Higgs mass bound because it generates an additional 
$F$-term contribution to the Higgs mass.  The tree-level bound now becomes
\beq
m_h^2 \leq M_Z^2\cos^22\beta + \lambda^2\,v^2\sin^22\beta.
\eeq
The value of the $\lambda$ coupling runs large in the UV,
and must be less than about $0.7$ if it is to remain 
perturbatively small up to $M_{GUT}$~\cite{nmssm}.  
Even so, the tree-level bound on the SM-like Higgs boson can be 
increased to about $110$~GeV for $\lambda$ as large as possible 
and $\tan\beta \simeq 2$, substantially ameliorating the Higgs mass
problem of the MSSM.

  Unfortunately, adding a singlet to the MSSM can create 
new problems.  In models of softly-broken supersymmetry derived from 
supergravity, both the scalar and $F$-term components of the singlet
develop large expectation values due to tadpole loops involving 
higher-dimensional, $M_{\rm{Pl}}$-suppressed operators.  These VEV's 
are proportional to positive powers of the cutoff scale and have the 
effect of destabilizing the electroweak scale~\cite{Bagger:1993ji}.  
This outcome may be avoided by introducing a symmetry under which 
the singlet is charged.  

  Such a symmetry may be discrete or continuous, global or gauged, but will
necessarily be broken when the would-be singlet field $S$ develops a 
VEV on the order of the electroweak scale.  Due to this breaking, 
ungauged symmetries are problematic: continuous global symmetries can generate 
troublesome axions~\cite{Miller:2003hm}; discrete symmetries often lead to 
cosmologically unacceptable domain walls in the early 
universe~\cite{Abel:1995wk}.  Moreover, attempts to get around these
problems often lead to new ones.  For example, domain walls can be avoided 
by weakly breaking the discrete symmetry with operators of 
dimension greater than four, but this reintroduces the danger of
destabilizing the electroweak scale by singlet tadpole loops.
In particular, within the next-to-minimal supersymmetric standard 
model~(NMSSM), a popular singlet model based on a discrete 
$\mathbb{Z}_3$ symmetry~\cite{nmssm}, it is not possible to avoid 
the domain wall problem while maintaining a stable gauge 
hierarchy~\cite{Abel:1995wk}.\footnote{Models based on $\mathbb{Z}^R_5$ 
and $\mathbb{Z}_7^R$ discrete $R$-symmetries do, however, appear to be 
viable~\cite{Abel:1996cr,n2mssm}.}
In light of these problems faced by global symmetries, we are led 
to consider gauge symmetries as a way to protect the $S$ field from large 
quantum corrections.

The simplest choice for such a symmetry, and the one we shall consider
in the present work, is a $U(1)_X$ gauge symmetry in addition to the 
$G_{SM}= SU(3)\times SU(2)\times U(1)_Y$ gauge symmetry of 
the MSSM~\cite{u1x,Erler:2000wu,Chamseddine:1995rs,Demir:2005ti}.
The choice of a $U(1)_X$ is also attractive because its lone gauge
boson will be massive after symmetry breaking.  Moreover, by protecting
the $S$ field with a gauge symmetry, there is an additional $D$-term
contribution to the Higgs mass which further increases the tree-level 
mass bound,
\bea
m_h^2 & \leq & M_Z^2\cos^22\beta + \lambda^2\,v^2\sin^22\beta \\
& & + 2g_x^2v^2(h_u\cos^2\beta + h_d\sin^2\beta)^2, \nonumber
\eea
where $g_x$ is the $U(1)_X$ gauge coupling, and $h_u$ and $h_d$ are
the charges of the $H_u$ and $H_d$ fields under this group.
   
  In extending the MSSM to include a gauged $U(1)_X$ and new particles, 
we would also like to preserve one of the particularly attractive 
features of the MSSM, namely the consistency of the model with 
grand unification.  Within the MSSM, the $\gsm$ gauge couplings unify 
at a high energy, of order $10^{16}$~GeV~\cite{unif}, where the 
deviation from exact unification is small enough that it could plausibly be 
explained by high-scale threshold corrections. Furthermore, the matter 
fields of the MSSM fit neatly into $\mathbf{\overline{5}}$ and $\mathbf{10}$ 
multiplets of $SU(5)$~\cite{Georgi:1974sy}.  These features 
suggest that the low-energy gauge structure follows from a grand-unified 
theory~(GUT).  However, in adding a $U(1)_X$ gauge symmetry, 
we must ensure that the corresponding anomalies vanish, and this generally 
entails adding additional fields that could potentially ruin the 
unification of gauge couplings.

  In the present work, we shall investigate whether it is possible
to extend the gauge group of the MSSM to include a gauged $U(1)_X$ under which
the $S$ field responsible for generating the $\mu$-term has a non-zero charge,
while maintaining gauge coupling unification.  
In doing so we will make the following assumptions:
\begin{enumerate}
\item All the terms present in the MSSM superpotential appear in the
superpotential of the extended model.
\item The $U(1)_X$ charges of the MSSM matter fields are family-universal.
(This is related to point 1, when assuming an unrestricted form of the Yukawa
couplings in the MSSM superpotential.)
\item The exotic matter needed to cancel the $U(1)_X$ anomalies
consists either of $G_{SM}$ singlets, or of complete $SU(5)$ multiplets
(with the usual $G_{SM}\subset SU(5)$ embedding).
\item The full set of exotic matter is vector-like in its $\gsm$ 
representation.
\end{enumerate}
The first condition ensures that the model reproduces the correct 
low-energy physics, while the second prevents the emergence 
of possibly dangerous flavor-mixing effects.  The third condition implies 
that the extended model will preserve gauge coupling unification.  
However, we will allow the states within each $SU(5)$ multiplet to have 
different $U(1)_X$ charges.  We will also demand that this unification 
occur in a perturbative regime, the implications of which will be 
discussed below.  The fourth condition ensures that the exotics
will not induce SM anomalies, or generate overly large corrections
to the precision electroweak observables.

  We will begin our analysis by classifying the different possibilities for 
the $\u1x$ symmetry, when acting on the MSSM fields, in Section~\ref{0:class}.
In Section~\ref{1:singlet}, we shall consider the addition of pure $\gsm$
singlets to the model.  With only these fields, we find that the 
$\gsm$ singlet field $S$ responsible for generating the effective 
$\mu$-term must also be a singlet under $U(1)_X$.
We are thus forced to look at more complicated extensions involving 
additional $\gsm$ non-singlet matter.  
The general implications of this 
matter on gauge coupling unification will be the subject of 
Section~\ref{2:unif}.  
In Section~\ref{3:new}, we will examine the 
implications of anomaly cancellation on the possible $U(1)_X$ charges 
of the MSSM fields as well as the exotics.  
In Section~\ref{4:ex}, we shall put our results to use by constructing 
a concrete model.  
Finally, Section~\ref{5:conc} is reserved for our conclusions.  
Some technical details and a list of $E_6$ charges are given in 
a pair of Appendices.

 Finally, we note that an investigation similar to the present one
has been performed in Ref.~\cite{Erler:2000wu}.  However, compared
to this work, our starting assumptions and therefore our final conclusions
are rather different.


\section{Classification of Possible $U(1)_X$'s\label{0:class}}

  Before adding exotic matter to the MSSM, let us first investigate
the most general action of a $\u1x$ symmetry on the fields of the MSSM.
We take the superpotential to be that of the MSSM up to the replacement 
of the $\mu$-term by a $G_{SM}$ singlet field $S$ and possible additional 
terms involving both the MSSM fields and an as yet unspecified set 
of exotics,
\bea
W & = &  y_u Q\ccdot H_u\,U^c 
- y_d Q\ccdot H_d\,D^c- y_e L\ccdot H_d E^c\label{1:w} \\
 & &  + \lambda\,S\,H_d\ccdot H_u + (exotics).\nonumber
\eea
The family-universal gauge charges under $G_{SM}\times U(1)_X$ 
are defined to be
\beq
\begin{array}{cccccc}
Q &=& (\mathbf{3}, \mathbf{2}, 1/6, q)&~~U^c &=& 
(\mathbf{\bar{3}},\mathbf{1},-2/3,u) \\
D^c &=& (\mathbf{\bar{3}},\mathbf{1},1/3,d)&~~L&=&
(\mathbf{1},\mathbf{2},-1/2,l) \nonumber \\
E^c &=& (\mathbf{1},\mathbf{1},1,e)&~~S&=&(\mathbf{1},\mathbf{1},0,s) 
\nonumber \\
H_u&=&(\mathbf{1},\mathbf{2},1/2,h_u)&~~ H_d&=&(\mathbf{1},\mathbf{2},-1/2,h_d)
\end{array}\label{1:fields}
\eeq
where $q, u, d, \ldots$ denote the  $U(1)_X$ charges.

  For the superpotential of Eq.~(\ref{1:w}) to be gauge invariant,
the $U(1)_X$ charges must sum to zero for each allowed operator:
\bea
q+u+h_u &=& 0\hspace{1cm}q+d+h_d = 0\label{1:gauge}\\
l+e+h_d&=& 0\hspace{1cm}s+h_u+h_d = 0\nnmb.
\eea
These equalities form a non-degenerate system of four equations
in eight variables, and allow us to solve for $q$, $u$, $l$, and $s$
in terms of $d$, $e$, $h_u$, and $h_d$.  Since there are four free variables,
it follows that the action of any $\u1x$ on the MSSM can be expressed 
as a linear combination of four independent basis $U(1)$'s.
\comment{
Two obvious candidates for such a basis are $U(1)_Y$ and $U(1)_{B-L}$.  
A convenient additional pair of $U(1)$'s that allow us to make 
complete this (non-orthogonal) basis are the $U(1)_{\psi}$ and 
$U(1)_{\chi}$ subgroups of $E_6$, with the charges of the MSSM and $S$ 
fields corresponding to an embedding in the $\mathbf{27}$ 
representation.}
One obvious candidate for this basis is $U(1)_Y$.  A convenient choice
for the rest of the basis is $U(1)_{B+L}$, and the $U(1)_{\chi}$
and $U(1)_{\psi}$ subgroups of $E_6$ with the charges of the MSSM and $S$
fields corresponding to an embedding in the $\mathbf{27}$ 
representation~\cite{Hewett:1988xc}.
These charges are listed in Appendix~\ref{app:e6}.
For a given $\u1x$ symmetry, specified by the set $\{d,e,h_u,h_d\}$,
the decomposition into the above-mentioned basis is given by
\bea
Q^i_X &=& \frac{2}{5}(-3d+e+2h_u-3h_d)\,Q^i_Y\label{xcharges}\\ 
&&+ \frac{1}{2}(-3d-e+h_u-3h_d)\,Q^i_{B+L}\nnmb\\
&& - \frac{1}{4}(h_u+h_d)2\sqrt{6}\,Q^i_{\psi}\nnmb\\ 
&&+ \frac{1}{20}(6d-2e+h_u+h_d)2\sqrt{10}\,Q^i_{\chi},\nnmb
\eea
where $Q^i_A$ is the charge of field $\phi_i$ under gauge group $U(1)_A$.
Notice that the singlet $S$ is protected and the standard MSSM $\mu$-term is 
forbidden by the $\u1x$ symmetry if and only if the $\u1x$ contains
a component of $U(1)_{\psi}$, since this is the only basis-$U(1)$ under 
which $S$ can be charged.  Because this conclusion follows 
from the gauge invariance of the required superpotential 
operators, it will continue to hold (for the MSSM fields) no matter what 
exotic fields are added to the model.


\section{Standard Model Exotics Required\label{1:singlet}}

The simplest way to realize an additional $U(1)_X$ gauge symmetry
while maintaining gauge coupling unification is to augment the 
MSSM by fields that are singlets under $G_{SM}$.  
Suppose that on top of the $S$ field that generates the $\mu$-term,
we include an additional set of $\gsm$ singlet fields.
Besides the conditions for gauge invariance of 
the needed superpotential operators, Eq.~(\ref{1:gauge}), the $\u1x$ charges
of all the fields are constrained by the requirement of anomaly cancellation.  
If the exotic matter consists only of $G_{SM}$ singlets, the pure SM 
anomalies vanish automatically.  The remaining anomaly conditions are due 
to the $\gsm\u1x$ mixed anomalies, $SU(3)^2U(1)_X$, $SU(2)^2U(1)_X$, 
$U(1)_Y^2U(1)_X$, and $U(1)_YU(1)_X^2$, as well as the 
gravitational-$\u1x$ and $\u1x^3$ anomalies.  The four mixed anomaly 
conditions depend only on the charges of $\gsm$ non-singlet fields, 
and allow only a highly restrictive choice of charge relations among 
the SM fields~\cite{Chamseddine:1995rs}.

  Consider, for instance, the constraint implied by the 
$SU(3)^2\,\u1x$ anomaly,
\beq
2q + u+ d = 0.
\eeq
Together with the relations in Eq.~(\ref{1:gauge}), this condition
implies that $h_u+h_d = 0$.  It follows that if we are to obtain a 
non-zero charge for the $S$ field, we must include colored exotics
in addition to $\gsm$ singlets.\footnote{
This conclusion may be avoided if we relax our assumptions on the 
form of the superpotential and family universality for the $\u1x$ charges.
See, for example, Ref.~\cite{Demir:2005ti}.}
This result may also be understood by the fact that when all exotic 
matter fields are $\gsm$ singlets, the only allowed, anomaly-free 
$U(1)_X$ gauge symmetries are linear combinations of $U(1)_Y$ and 
$U(1)_{B-L}$ when acting on the MSSM~\cite{Chamseddine:1995rs}.

\section{Exotics and Unification Relations\label{2:unif}}

The problem with colored exotics, which are necessary if the $S$ field 
is to be protected by the $\u1x$ symmetry, is that they interfere with the
unification of gauge couplings that occurs in the MSSM.  To avoid
disturbing the unification relations, the exotic matter should shift the 
three $\gsm$ gauge coupling $\beta$-functions by an equal amount.  
This is automatic provided the new matter comes in complete $SU(5)$ multiplets,
and we will focus exclusively on this possibility in the present work.

  There is also a good cosmological reason for this restriction.
Exotic quarks should have the same electromagnetic charges as their SM 
counterparts, otherwise they would be stable, and the expected thermal relic 
abundance of a stable heavy quark would violate the experimental bounds 
on the anomalously heavy baryons they would produce~\cite{Nardi:1990ku}.
If we restrict ourselves to color-triplet exotics that have the same
$SU(3)\times U(1)_{em}$ quantum numbers as their MSSM counterparts, 
the two smallest sets of exotics that shift the $\gsm$ $\beta$-functions 
by an equal amount have the $\gsm$ quantum numbers of the 
$\mathbf{5}$ and $\mathbf{10}$ multiplets up to the signs of the
hypercharges of the lepton-like states.  This sign ambiguity is inconsequential
if we further require that the exotics come in vector-like sets with respect
to their $\gsm$ quantum numbers.  In this sense, exotics in the form
of $\vfive$ or $\vten$ multiplets are the minimal vectorial possibilities 
consistent with unification.

  In order to obtain constraints on which and how many 
$SU(5)$ multiplets are sensible to consider, we review here some 
general features of supersymmetric gauge coupling unification.
The scale dependence of the gauge
couplings at one-loop is described by 
\beq
\frac{d\alpha_i}{dt} = -\frac{b_i}{2\pi}\,\alpha_i^2,
\label{2:beta2}
\eeq
where $\alpha_i = g_i^2/4\pi$ and $t = \ln(Q/M_Z)$.
For an $N=1$ supersymmetric gauge theory, the coefficient $b_i$ is given by
\beq
b_i = 3\,C_2(G_i) - \sum_{r_i} S_2(r_i),
\eeq
where $C_2(G_i)$ is the quadratic Casimir invariant for the 
adjoint of the \emph{i-th} gauge group, and $S_2(r_i)$ is 
the trace invariant for the matter representation $r_i$ of $G_i$.
We take $g_i$ to be normalized according to the usual
embedding of $G_{SM}$ in $SU(5)$, so that the $g_1$ coupling is related
to the $U(1)_Y$ coupling by $g_1 = \sqrt{5/3}\,g_Y$, and $Q_1 = \sqrt{3/5}\,Y$.
The $\beta$-function coefficients $b_i$ for the MSSM are therefore
$(b_1, b_2, b_3) = (-33/5, -1, 3)$.

  The success of unification within the MSSM is usually illustrated
by setting the unification point $(M_G, \alpha_G)$ by the condition
$\alpha_1(M_G) = \alpha_2(M_G) := \alpha_G$, and using this point
to generate a prediction for $\alpha_3(M_Z)$.  Taking as inputs the
values
$\alpha_1^{-1}(M_Z) \simeq 59.1$ and $\alpha_2^{-1}(M_Z) \simeq 29.4$,
and using the solution to Eq.~(\ref{2:beta2}),
\beq
\alpha_i^{-1}(Q) = \alpha_i^{-1}(M_Z) + \frac{b_i}{2\pi}t,
\eeq
we obtain
$\alpha_G^{-1} \simeq 24.1$, $M_G \simeq 2.7\times 
10^{16}\,\mbox{GeV}$, and $\alpha^{-1}_3(M_Z) \simeq 8.2$.
This value for $\alpha^{-1}_3(M_Z)$ is near 
the measured value of about 8.7, and is compatible
with unification when higher-order corrections and 
reasonable high-scale thresholds are taken into account~\cite{unif}.

  To preserve the prediction of unification, any additional
matter in the model should generate an equal shift in each of the 
$b_i$ coefficients.  This occurs automatically if the new matter
comes in complete $SU(5)$ multiplets,
\bea
\Delta b_i & = & - \frac{1}{2}\left[(N_5+N_{\bar{5}}) 
+ 3(N_{10}+N_{\overline{10}})
\right. \\
& & ~~~\left. + 7(N_{15}+N_{\overline{15}}) 
+ 10\,N_{24} + \ldots~\right], \nonumber
\eea
where $i=1,2,3$, and $N_5$ is the number of $\mathbf{5}$'s and so on.  
Shifting the $b_i$ in this way does not change the unification scale, 
but it does increase the value of the unified gauge coupling,
\bea
\alpha_G^{-1} & \to & \alpha_G^{-1} + \frac{\Delta b}{b_2-b_1}
\left[\alpha^{-1}_1(M_Z) - \alpha^{-1}_2(M_Z)\right] \\
& \simeq &  \alpha^{-1}_G + (5.3)\Delta b. \nonumber
\eea
We will take $\alpha_G^{-1} > 1$ as the condition for 
perturbative unification which puts an upper limit on the number of
new $SU(5)$ multiplets; 
$N_5 \leq 8$, $N_{10}\leq 2$, $N_{15}\leq 1$,
and $N_r = 0$ for any of the higher-dimensional representations.
Although it is possible to have a single $\mathbf{15}$ and maintain 
perturbative unification, it is not possible to do so while canceling the 
$G_{SM}$ gauge anomalies, which are all proportional to
\beq
(N_5-N_{\overline{5}}) + (N_{10}-N_{\overline{10}}) 
+ 9(N_{15}-N_{\overline{15}}) + \ldots.
\eeq
Thus, we need only consider $\mathbf{5}$'s, $\mathbf{10}$'s, 
and their conjugates.
 
  We may also reverse the argument in the previous paragraph
to constrain the possibility of $\gsm$ gauge coupling unification 
using the number and charges of the TeV-scale exotics that are observed
in experiment.  For instance, the discovery of more than two $\mathbf{10}$'s or
eight $\mathbf{5}$'s (or more precisely, of particles having these 
quantum numbers) would preclude perturbative unification.  
This constraint is particularly relevant for the $\u1x$ 
gauge symmetry because, as we shall see, it is 
frequently the case that a large number of exotics, charged under $\u1x$,
are required to cancel the gauge anomalies.  In principle, colliders could 
extract the mass of the $Z'$ gauge boson as well as the couplings 
$g_x\,Q_i$, where $g_x$ is the $\u1x$ gauge coupling 
and $Q_i$ is the charge of the $i$-th species, for all the SM states 
and exotics that are light enough to be probed.
With this knowledge, we could test whether the known charges are
consistent with a perturbative theory up to the GUT scale.  Requiring that no
Landau pole develop below the GUT scale (at one loop) puts a limit on the
gauge coupling and charges of the low-scale states:
\beq
\sum d_i(g_xQ_i)^2 \lsim \frac{8\pi^2}{\ln (M_{\rm GUT}/M_Z)}\simeq 2.4,
\eeq
where $d_i$ is the dimensionality of the representation of the 
$i$-th species.  One must be careful in model building that the 
$g_xQ_i$ charges are not too large, otherwise the $U(1)_X$ gauge
coupling will go strong and mix with the SM gauge couplings 
at higher loop order, destroying predictable, perturbative gauge 
coupling unification.

\section{Non-Singlet Exotics\label{3:new}}

  As shown in Section~\ref{1:singlet}, exotic colored matter is needed to
cancel the $U(1)_X$ anomalies and to give the $S$ field a non-zero $U(1)_X$
charge.  In this section, we attempt to accomplish this by adding new matter
in the form of complete $SU(5)$ multiplets.  

To begin, we will assume that the exotic $SU(5)$ multiplets have
universal $\u1x$ charges, as would be expected if these multiplets were
indeed derived from $SU(5)\times\u1x$.  On the other hand, for the
sake of generality, we will not immediately impose such a condition on 
the charges of the MSSM matter fields, even though they do fill out
complete $SU(5)$ multiplets.

  We denote the $U(1)_X$ charges of the exotic $SU(5)$ multiplets by 
$D_a$ and $\overline{D}_a$, $a=1,2,\ldots$ for the ${\mathbf{5}}$'s 
and the $\overline{\mathbf{5}}$'s, and $Q_b$ and $\overline{Q}_b$ for the 
$\mathbf{10}$'s and $\overline{\mathbf{10}}$'s.  
With the exotics, the conditions for the gauge invariance of the 
operators in the superpotential are still given by Eq.~(\ref{1:gauge}),
while the cancellation of $SU(3)^2U(1)_X$, $SU(2)^2U(1)_X$ 
and $U(1)_Y^2U(1)_X$ anomalies implies
\bea
n_g(2q+u+d) + M &=&0,\label{3:su3x}\\
n_g(q+u+l+e) + h_+ + M &=& 0,\label{3:su2x}\nonumber\\
n_g(q+8u+2d+3l+6e) + 3h_+ + {5}M &=& 0.\label{3:u1x}\nonumber
\eea
where $M = \sum_a (D_a + \overline{D}_a) + 3\sum_b (Q_b+\overline{Q}_b)$,
$h_+ = (h_u+h_d)$, and $n_g = 3$ is the number of MSSM generations.
Note that these equations would be unchanged had we also included
an arbitrary number of $\gsm$ singlets.

  It is not hard to see that taken together, Eq.~(\ref{1:gauge})
and Eq.~(\ref{3:su3x}) imply that both $s\, (=-h_u-h_d)$ and $M$ 
vanish. Therefore, it is not possible to protect the $S$ field by 
giving it
a $\u1x$ charge if the $\gsm$-charged exotics, in the form of $SU(5)$
multiplets, have a universal $\u1x$ charge within each multiplet.
This result holds for larger $SU(5)$ multiplets as well.

\subsection{Non-Universal $\u1x$ Exotics}

  The main result of the above analysis is that the exotic $SU(5)$ 
multiplets cannot have universal $\u1x$ charges within each multiplet
if $S$ is to be charged under this gauge symmetry.  On the other
hand, if we allow for non-universal $\u1x$ charges within the exotic
$SU(5)$ multiplets, we find that it is possible to obtain a non-zero 
$\u1x$ charge for the $S$ field.  A simple extension of the MSSM that 
illustrates this result consists of a single extra $\vfive$,
\bea
\mathbf{\bar{5}} &=& D_1^c\oplus L_1 = 
(\mathbf{\bar{3}},\mathbf{1},1/3,\overline{D}_1)\oplus(\mathbf{1},
\mathbf{2},-1/2,\overline{L}_1),\nnmb\\ 
\mathbf{5} &=& D_1\oplus L_1^c = 
(\mathbf{3},\mathbf{1},-1/3,{D}_1)\oplus(\mathbf{1},
\mathbf{2},1/2,L_1),\nnmb
\eea
and the $\gsm$ singlets
\bea
S =  (\mathbf{1},\mathbf{1},0,s),&~&A = (\mathbf{1},\mathbf{1},0,a),\\
B = (\mathbf{1},\mathbf{1},0,b)&~&Z_m = 
(\mathbf{1},\mathbf{1},0,z_m).\nonumber
\eea
As before, the $S$ field will be responsible for generating the $\mu$-term, 
while $A$ and $B$ are new Higgs fields included to ensure that the  
$\vfive$ exotics, $D_1^c$, $D_1$, $L_1$, and $L_1^c$, obtain large 
tree-level masses.  The $Z_m$ represent any other $\gsm$ singlets that 
might be present in the model.

  If we allow no exotic fields besides the $\vfive$, $S$, $A$, and $B$,  
(\textit{i.e.} we exclude all the $Z_m$)
we find that there exist solutions that satisfy the necessary constraints,
but that all of these solutions require charges that are irrational relative 
to each other.  The reason for this is that the condition implied
by the $\u1x^3$ anomaly is cubic in the charges.  We skip the details of 
this derivation since such relative irrational solutions appear to us to 
be problematic when trying to embed them into a grand unified framework.

Instead, we focus on the full complement of exotic states, 
including additional $Z_m$ singlets.
The conditions for the cancellation of anomalies involving $\gsm$ subgroups, 
such as $SU(3)^2\u1x$, depend only on the charges of the MSSM and $\vfive$ 
fields.  On the other hand, the conditions required for the cancellation 
of gravitational-$\u1x$ and $\u1x^3$ anomalies depend on the charges of 
all the fields, including the $\gsm$ singlets $S$, $A$, $B$, and $Z_m$.  
The latter two conditions turn out to present no significant constraint 
at all; for any solution to the singlet-independent gauge and anomaly 
conditions with rational $U(1)_X$ charges, it is always possible to 
satisfy the remaining two $\gsm$ singlet-dependent equations by a 
judicious choice of rational $\u1x$ charges for the $Z_m$ fields.  
Our proof is given in Appendix~\ref{app:ratio}.
Note, however, that there does not always exist a rational solution to 
the singlet-independent equations on account of the $U(1)_YU(1)_X^2$
condition which is quadratic in the $U(1)_X$ charges.  Even so, for all 
the cases studied here, in which the $\gsm$ non-singlet
exotics consist of complete $SU(5)$ multiplets, it has been possible
to find a rational solution to the singlet-independent conditions.


\subsection{$SU(5)$-Compatible Charge Assignments}

 Since the quantum numbers of the MSSM matter fields fill out
complete $SU(5)$ multiplets, it is natural to arrange their $U(1)_X$
charges to be consistent with an embedding in $SU(5)\times U(1)_X$.  
The necessary conditions for this are 
\beq
q = u = e, ~~~~~\mbox{and}~~~l = d.\label{2:su5charge}
\eeq
Unfortunately, as we found above, it is not possible to apply
this condition to the $SU(5)$ exotics without forcing $s=0$
as well.

For the simple case of one set of $\vfive$ exotics, the singlet-independent 
gauge and anomaly conditions imply the following relations between 
the $\u1x$ charges:
\bea
q &=& -h_-/4 - L_+/4(n_g-1)\label{2:su5}\\
l &=& 3h_-/4 - L_+/4(n_g-1)\nnmb\\
h_+ &=& L_+/(n_g-1)\nnmb\\
D_+ &=& n_gL_+/(n_g-1)\nnmb\\
D_- &=& -h_-/n_g - (n_g-1)L_-/n_g,\nnmb
\eea
where $h_\pm =h_u\pm h_d$, $D_\pm=\overline{D}_1\pm D_1$, 
$L_\pm=\overline{L}_1\pm L_1$, and $n_g=3$ is the number of generations.
For the MSSM fields, we see that their charges depend only
on the two parameters $h_-$ and $L_+$.  Notice as well that it is
not possible to have $D_+ = L_+$ unless both they and $h_+$ vanish,
illustrating the necessity of non-universal charges for the exotics.\footnote{
If we include $n_g$ sets of $\vfive$'s, we find the same relationship 
between $D_+$ and $L_+$ as in Eq.~(\ref{2:su5}).}

  The relationships of Eq.~(\ref{2:su5}) show that when the 
$SU(5)$ condition of Eq.~(\ref{2:su5charge}) is imposed, the $\u1x$ 
charges of the MSSM fields can be expressed in terms of just two parameters,
$L_+$ and $h_-$.  Indeed, these charges are such that the $U(1)$'s 
generated by $h_-$ and $L_+$ coincide with the $U(1)_{\chi}$ and $U(1)_{\psi}$ 
subgroups of $E_6$, respectively, with the charges of the MSSM fields 
corresponding to their counterparts in the $\mathbf{27}$,
as listed in Appendix~\ref{app:e6}.  
This is in accord with what we found in Section~\ref{0:class}.  
In the present case, the $SU(5)$ condition of Eq.~(\ref{2:su5charge}) forces 
the hypercharge and $(B\!+\!L)$ components of $\u1x$ to vanish, leaving 
behind only the two $E_6$ subgroups, $U(1)_\chi$ and $U(1)_\psi$.  
Unfortunately, the charges of the $\vfive$ exotics do not follow 
this pattern, which makes it difficult to establish a GUT interpretation.


\subsection{$SO(10)$-Compatible Charge Assignments}

 The MSSM matter fields, when augmented by three generations of 
singlet neutrinos, $N^c = (1,1,0,n)$, fill out complete
$\mathbf{16}$'s of $SO(10)$.  It is therefore also natural to
arrange the $U(1)_X$ charges to be be consistent with such an embedding,
implying
\beq
q = u = d = l = e.\label{2:so10charge}
\eeq
This condition is a further restriction on the one given 
in Eq.~(\ref{2:su5charge}) and imposes an even stronger constraint
on the $U(1)_X$ charges of the MSSM fields.  By comparing with 
Appendix~\ref{app:e6}, we see that Eq.~(\ref{2:so10charge}) projects out the
$U(1)_{\chi}$ component of $\u1x$, and implies that the action of 
$U(1)_X$ is just that of $U(1)_{\psi}\in E_6$, with the MSSM field 
charges corresponding to those of the $\mathbf{27}$ up to normalization.  
Furthermore, this condition always forces $h_- = 0$.

To illustrate this result, consider the relationships between
the charges with an exotic $\vfive$.  
Applying Eq.~(\ref{2:so10charge}) to Eqs.~(\ref{2:su5}), we find
\bea
q &=& - L_+/4(n_g-1)\label{2:so10}\\
h_+ &=& L_+/(n_g-1)\nnmb\\
h_- &=& 0\nnmb \\
D_+ &=& n_gL_+/(n_g-1)\nnmb\\
D_- &=& - (n_g-1)L_-/n_g.\nnmb
\eea
The charges of $q$, $h_+$, and $h_-$ coincide precisely with their
values under $U(1)_{\psi}$, up to normalization.
It is fortunate that our surviving $U(1)_\psi$ happens to also protect
the $\mu$-term. This nice feature is implicit in the example model
we present in the next section.


\section{Example Model\label{4:ex}}

  In order to construct a concrete example, and to illustrate the technique
outlined in Appendix~\ref{app:ratio}, we investigate the $U(1)_X$-augmented
MSSM subject to the condition of Eq.~(\ref{2:so10charge}) with a single
$\vfive$ as the only exotics charged under $\gsm$.
The $\gsm$ singlets we include are the $S$ field which generates the 
$\mu$-term, the $A$ and $B$ fields which give mass to the $\vfive$ 
exotics, and some number of $Z_m$ such that the $U(1)_X$ charges end up 
being rational.  For simplicity, we will set $L_- = 0$.

  With the singlet content described above, the conditions for the 
cancellation of the gravitational-$\u1x$ and $\u1x^3$ anomalies reduce to
\bea
\sum_m z_m &=& \frac{9}{8}\,L_+ = \alpha, \\
\sum_m z_m^3 &=& \frac{765}{512}\,L_+^3=\beta\nnmb
\eea
Without loss of generality, we may set $L_+ = -b = -8$.
Setting $z_0 = -9$, and noting that $(\beta-\alpha^3) = -36$,
a possible choice for the additional singlet charges is therefore 
$\{z_0,z_1,z_2,z_3\} = \{-9, -4, 3, 1\}$. With this choice, 
the $U(1)_X$ charges are
\beq
\begin{array}{lcl}
q=u=d=l=e = 1&~~&s = 4\\
(z_0,z_1,z_2,z_3) = (-9,-4,3,1)&&h_u=h_d = -2\\
\overline{D}_1={D}_1 = -6&&a=12\\
\overline{L}_1={L}_1 = -4&&b = 8
\end{array}
\eeq

  For these charges, the list of gauge-invariant supersymmetric 
operators of lowest dimension is:\\

$d=2$ Superpotential
\beq
\begin{array}{ccccc}
SZ_1
\end{array}
\eeq
$d=3$ Superpotential
\beq
\begin{array}{ccccc}
QU^cH_u&QD^cH_d&LE^cH_d&SH_uH_d&\\
AD_1D_1^c&BL_1\overline{L}_1&Z_1Z_2Z_3&Z_1Z_1B&BZ_0Z_3\\
LH_uZ_3&LL_1^cZ_2
\end{array}
\eeq
$d=4$ Superpotential
\beq
\begin{array}{cccc}
LL_1^cAZ_0&E^cH_dH_uZ_2&H_uH_dBZ_1&H_uH_dZ_2Z_3 \\
H_uL_1Z_2Z_2 & H_dL_1^cZ_2Z_2 & SSL_1{L}_1^c& SSZ_0Z_3 \\
SSZ_1Z_1 & SD_1D_1^cB & L_1{L}_1^cAZ_1 & AZ_0Z_1Z_3 \\
AZ_1Z_1Z_1 & Z_0Z_2Z_2Z_2 & &
\end{array}
\eeq
$d=3$ K\"ahler Potential ($+ h.c.$)
\beq
\begin{array}{ccccc}
L^{\dagger}H_dZ_2&S^{\dagger}BZ_1&S^{\dagger}Z_2Z_3&A^{\dagger}SB&\\
B^{\dagger}SS&B^{\dagger}AZ_1&Z_1^{\dagger}H_uH_d&Z_2^{\dagger}AZ_0&
\end{array}
\eeq

  Of these operators, the dangerous ones are $SZ_1$ and $LH_uZ_3$.
The first, if allowed, would reintroduce a dimensionful coupling
into the superpotential which must be of order the electroweak scale
for the VEV of $S$ (and consequently $\mu_{eff} = \lambda\left<S\right>$)
to be of the right size.  This is precisely the sort of \emph{$\mu$ problem}
we set out to avoid in the first place.  
The second operator would induce a mixing between 
the leptons and the Higgsinos of the MSSM if $Z_3$ develops a VEV, possibly 
leading to overly large lepton masses.  Similarly, the operator 
$LL_1^cZ_2$ can also be problematic since it induces a mixing between 
the MSSM leptons and their exotic counterparts, but it may also be useful
in that it facilitates the decay of the heavy leptons.  
All three of these operators can be forbidden by
imposing an $R$-parity symmetry under which
\bea
Q \hspace{-0.15cm}& = &\hspace{-0.15cm} U^c=D^c=L=E^c=  -1 \\
D_1\hspace{-0.15cm} & = & \hspace{-0.15cm}\overline{D}_1=L_1=\overline{L}_1=Z_1 
= Z_2\:=\: -1,\nnmb\\
H_u\hspace{-0.15cm} & = &\hspace{-0.15cm}H_d=S=A=B=Z_0=Z_3 \:=\: +1. \nonumber
\eea
This set of charges still allows terms such as $Z_1Z_1B$, $Z_1Z_2Z_3$,
and $BZ_0Z_3$ which help to ensure that the fermionic components of the
singlets other than the $A$ get large masses provided all the $R$-even $Z_m$ 
fields develop VEV's.  Note also that, even without imposing an $R$-parity, 
the supersymmetric terms that could generate dimension five Lagrangian 
operators leading to proton decay are forbidden by the $U(1)_X$.

  As it stands, the model still has a couple of potential problems.
The first is that the $A$ field does not get a contribution to its 
mass from any of the renormalizable superpotential operators. 
Instead, the fermion component of this field only receives a mass 
through its mixing with the $\u1x$ gaugino of order $g_x\left<\phi_A\right>$.  
If the low-energy $g_x$ coupling is very weak, as might be required 
to preserve perturbative unification in the presence of a large number 
of singlets, there will be a very light, mostly singlet fermion in the 
spectrum.  If this state is the LSP, which is quite likely for a light state, 
it will be stable on account of $R$-parity and could overclose 
the universe due to the feebleness of its couplings.  Thus, it would be 
reassuring if there was also a superpotential contribution to the 
$A$ fermion mass.\footnote{Another reason to desire this reassurance is that
if the model contained a second $R$-even field with no superpotential 
mass term, the neutralino mass matrix would have a zero eigenvalue 
for any value of the gauge coupling $g_x$.}

  The second problem is that the mixing between the exotic quarks 
from the $\vfive$ and those of the MSSM is highly suppressed, 
being relegated to non-renormalizable operators of dimension greater 
than five.
Such mixing induces a flavor-violating coupling between
the exotic quarks, the light quarks, and the $W$ gauge boson.
If these operators are suppressed by powers of $M_{\rm Pl}$
or $M_{GUT}$, the heavy exotic quarks will be cosmologically stable.
As discussed in Section~\ref{2:unif}, this leads to a relic abundance
of heavy baryons well above the experimental bounds.
This conclusion regarding the stability of heavy baryons
holds without question if the mixing arises from operators of dimension six 
or higher, as we find here.  For dimension-five mixing, 
suppressed by a single power of $M_{\rm Pl}$, the lifetime of the heavy 
baryons is of order $10^4$~s, long enough that they could potentially 
interfere with the predictions of Big Bang nucleosynthesis~(BBN) 
when they decay~\cite{Kawasaki:2004qu}.
However, the relic abundances for the heavy baryons found in 
Ref.~\cite{Nardi:1990ku} are low enough that the net effect of their 
decays on the light element abundances is acceptably small.

  One way to resolve both problems is to include more $\gsm$ 
singlet fields.  To generate a superpotential contribution to the mass 
of the $A$ field, we will add one set of $R$-parity even fields having 
$\u1x$ charges $\{a,5a,-6a\}$, and a second with charges 
$\{-2a,-3a,5a\}$.  As discussed in Appendix~\ref{app:eng}, these sets do not 
contribute to the anomalies and allow for cubic operators in the 
superpotential that lead to masses for the $A$ and all of the additional
fields provided some of these fields develop VEV's.
These sets are also chiral, in that they do not allow any explicit 
bilinear couplings between themselves.  In the present case, 
the additional fields do not induce any new bilinear couplings with
the other fields in the model either.
In the same way, we will induce a mixing between MSSM down quarks and
their heavy counterparts, allowing the heavy quarks to decay via the $W$
gauge bosons, by introducing a SM singlet field having $\u1x$
charge equal to $-(d+\overline{D}_1) = -(q+D_1)=5$.  
If this new field is embedded in the larger set of fields having charges 
$\{5,25,-30;-10,-15,25\}$, no anomalies will be generated, all the 
fields within the set will obtain superpotential contributions to 
their masses, and no new bilinear couplings will be induced.  
A general method for this sort of ``singlet engineering'' is 
given in Appendix~\ref{app:eng}.


\section{Conclusions\label{5:conc}}

One of the most appealing ways to explain the $\mu$-term in supersymmetry
is to promote it to a condensing scalar field charged under a new
$U(1)_X$ gauge symmetry.  Under the conservative assumptions 
stated in the introduction, we have shown that this idea leads to 
several interesting implications:
\begin{enumerate}
\item The most general $\u1x$ symmetry, when acting on the fields of the MSSM, 
is a linear combination of $U(1)_Y$, $U(1)_{B+L}$, 
and the $U(1)_{\psi}$ and $U(1)_{\chi}$ subgroups of $E_6$.  Of these basis 
$U(1)$'s, only $U(1)_{\psi}$ is able to forbid the $\mu$-term and protect the 
singlet $S$ that replaces it.  For an embedding of the MSSM in 
$SU(5)\times\u1x$, only the $E_6$ subgroups are allowed.  
For an embedding in $SO(10)\times\u1x$, only $U(1)_{\psi}$ is possible.
\item Anomaly cancellation requires the introduction of 
exotics charged 
under $\gsm$ to cancel all mixed anomalies in the theory.  
Such SM exotics can only be dismissed if the effective 
$\mu$-term is not charged under the $U(1)_X$.
\item Adding complete multiplets of $SU(5)$ according to their SM charges, 
which is required for ``automatic gauge coupling unification", necessitates 
assigning different $U(1)_X$ charges to the SM-like component states 
within each exotic multiplet.  
\item Solutions with rational $U(1)_X$ charges for all SM-charged states 
generally require a large set of $Z_m$ singlet states; nevertheless, 
a solution to the singlet-dependent anomaly equations from these states 
$Z_m$ is guaranteed, and we have shown an algorithm to obtain that solution.
\end{enumerate}

It is apparent from this study that if we want the $\mu$-term to 
originate from the spontaneous breaking of a $U(1)_X$ theory, it is 
difficult to accommodate the resulting set of states in the model within 
an $SU(5)\times U(1)_X$ theory.  One way to embed it in this fashion, 
however, would be to assume that the various SM fields that can be classified
as coming from $\vfive$ are really pieced together from parts of 
a larger set of $\vfive$ representations, each with its own $U(1)_X$ charge.  
Such representations arise naturally if the $SU(5)$ group
is derived from a larger group such as $SO(10)$ or $E_6$,
and the splitting of the multiplets may be achieved by way of an
orbifold compactification of a fifth dimension~\cite{Hall:2001pg}.  
On the other hand, from this point of view there is no obvious reason
why the low-energy spectrum of $\gsm$-charged exotics should consist
of complete $SU(5)$ multiplets.  Therefore the fact that we obtain 
gauge coupling unification taking into account only the fields of the 
MSSM would appear to be a fortuitous accident.  

  Another attractive possibility within this scenario would be to
embed the entire gauge structure, including the $U(1)_X$ gauge group, 
within $E_6$~\cite{Hewett:1988xc,E6 papers}.  
This is particularly compelling since we have shown that abelian
subgroups of $E_6$ are natural candidates for the $\u1x$.
They are the only possible candidates when $SU(5)$ or $SO(10)$ 
conditions are imposed on the charges of the MSSM fields.
A natural special case along these lines, studied recently in 
Ref.~\cite{King:2005jy}, is to embed the MSSM fields in three
$\mathbf{27}$'s of $E_6$, and to identify the $\u1x$ with an unbroken
linear combination of $U(1)_{\psi}$ and $U(1)_{\chi}$.  To preserve
gauge unification, an additional pair of Higgs-like doublets is 
needed, and this pair must be vectorial in its $\u1x$ charge to avoid
generating anomalies.  Such a model can solve the $\mu$ problem
of the MSSM and increase the Higgs boson mass~\cite{King:2005jy};
however, there is also a new (arguably less severe) $\mu$ problem 
associated with unification since the vectorial pair of Higgs-like doublets 
must survive to low energies.  Our results suggest that to preserve 
unification while avoiding such $\mu$ problems altogether is a difficult task.
As in the $SU(5)\times\u1x$ case, the embedding of the requisite charged 
exotics within $E_6$, and the splitting of their corresponding 
multiplets, presents an acute model-building challenge.

\comment{
A recent concrete example of this sort of model has been given 
in Ref.~\cite{King:2005jy}, in which the MSSM fields are embedded in 
three $\mathbf{27}$'s, and the $\u1x$ gauge symmetry is a obtained 
as a particular linear combination of $U(1)_{\psi}$ and $U(1)_{\chi}$.  
An additional pair of Higgs-like doublets are included to 
preserve gauge coupling unification.  This model solves the $\mu$ problem 
of the MSSM and helps to increase the mass of the lightest Higgs boson.
However, to preserve unification, the pair of doublets must be light,
and to avoid $\u1x$ anomalies, the pair must also be vector-like.  
This leads to a new (but arguably less severe) $\mu$ problem associated 
with gauge unification.  Our results suggest that to avoid such 
problems altogether is a complicated task.
As in the $SU(5)\times\u1x$ case, the embedding of the requisite charged 
exotics within $E_6$, and the splitting of their corresponding 
multiplets, presents an acute model-building challenge.}

  It is our view that a $\mu$-term originating from a charged singlet that 
breaks a $U(1)_X$ symmetry implies a degree of tension with 
supersymmetric gauge coupling unification.  The picture of grand 
unification in the presence of such a symmetry is necessarily much more 
complicated than in the MSSM.  In this paper we have developed some tools for 
constructing models of this type, and we have exhibited a particular
self-consistent example. We have also pointed out some of the peculiarities 
required of these models that may overwhelm their attractive features.  
In any event, it should be possible for experiment to look for 
the charged exotics, the singlets, or the $Z'$ gauge boson associated 
with this explanation of the $\mu$-term.  If confirmed, we may need to 
rethink unification.


\appendix

\section{Singlet Engineering\label{app:a}}

\subsection{Solutions to non-SM Anomalies\label{app:ratio}}

  Suppose the set of charges $\{q_i\}$ of the $\gsm$-charged fields,
with all $q_i$ rational, is a solution of the singlet-independent conditions.
For any such solution, it is possible to uniformly rescale the 
charges such that they are integers.  We will assume this has been done.
Denoting the $\gsm$ singlet charges under $U(1)_X$ as $z_m$, $m=0,1,\ldots N$,
the $ggU(1)_X$ and $U(1)_X^3$ anomaly equations have the form
\beq
\sum_{m=0}^N z_m = \alpha,~~~~~
\sum_{m=0}^N z_m^3 = \beta,\nnmb
\eeq
where $\alpha$ and $\beta$ are linear and cubic polynomials in
the $q_i$, respectively, with integer coefficients.  It follows
that $\alpha$ and $\beta$ are integers as well.
Now suppose we choose $z_0 = \alpha$, so that 
\beq
\sum_{m=1}^N z_m = 0,~~~~~
\sum_{m=1}^N z_m^3 = \beta-\alpha^3.\nnmb
\eeq
The remaining $z_m$ charges thus form an \emph{N-th partition of zero}.
Here is where a helpful theorem enters: The cubic sum $(\sum z^3_m)$ of any 
integer partition of zero $(\sum z_m=0)$ is a multiple of six.  Using this 
factor of six, we have many options to satisfy the cubic equation, including
\bea
36(\beta-\alpha^3)~{\rm sets} & {\rm of} & z_m=\left\{\frac{2}{6},-\frac{1}{6},-\frac{1}{6}\right\} \nonumber\\
6(\beta-\alpha^3)~{\rm sets} & {\rm of} & z_m=\left\{\frac{4}{6},-\frac{3}{6},-\frac{1}{6}\right\} \\
\beta-\alpha^3~{\rm sets} & {\rm of} & z_m=\left\{\frac{7}{6},-\frac{5}{6},-\frac{1}{6},-\frac{1}{6}\right\} \nonumber
\eea
The same solutions apply if $(\beta-\alpha^3)<0$,  except the above equations 
are multiplied by $-1$.  If $(\beta-\alpha^3)$ happens to be a multiple of six,
a solution is always possible with integer partitions of zero, the simplest of 
which is $(\beta-\alpha^3)/6$ copies of $z_m=\{2,-1,-1\}$.

Thus, there is always a rational solution to the singlet-only equations as 
claimed.  Other techniques for solving the anomaly equations can be found 
in~\cite{Batra:2005rh}.

\subsection{Constructing Masses without $\mu$-Terms\label{app:eng}}

  In formulating models involving an additional gauged $\u1x$, 
it is often the case that for a given anomaly-free set of fields, 
the gauge symmetries do not allow certain operators that are necessary
to obtain a realistic phenomenology.  The most important of these operators 
are usually bilinear in the fields, to generate masses for instance.
Two examples of this sort were encountered in the model studied in 
Section~\ref{4:ex}.  In both examples, a solution was obtained by adding
additional  singlet fields\footnote{In this appendix, we refer to 
fields that are uncharged under $\gsm$ as singlets, even though we will 
implicitly demand that they have a non-zero $\u1x$ charge.}, 
and an effectively bilinear operator was 
then constructed by choosing the charge of one of the singlets
to allow a trilinear operator consisting of the singlet and the 
desired bilinear product, and then arranging for the singlet to obtain a VEV.  
Two potential problems with such a construction are that the new singlets will 
contribute to the gravitational-$\u1x$ and $\u1x^3$ anomalies, 
and there is no guarantee that they will all obtain large masses.
We also wish to add singlets that are chiral in the sense that their
charges do not allow bilinear superpotential couplings that would
reintroduce a $\mu$ problem to the model.
In this appendix, we show how to build effectively bilinear operators
in this way by adding singlets, but without generating any of the 
above-mentioned problems.

  If the new singlets are to avoid contributing to the 
gravitational-$\u1x$ anomaly, they must form an \emph{N-th partition of zero},
as we discussed in Appendix~\ref{app:ratio}.  The smallest chiral
set with this feature consists of three fields.  Unfortunately
a single (chiral) triplet of this sort will necessarily contribute to the 
$\u1x^3$ anomaly by an amount equal to the sum of the cubes of its charges.
Thus, to avoid reintroducing this anomaly, several triplets are needed.
Two chiral triplet sets that work in this regard are
\bea
X_a &=& \left\{1,5,-6\right\},~~~Y_b = \left\{-2,-3,5\right\};
\label{2:sets}\\
{X_a}' &=& \left\{2,6,-8\right\},~~~{Y^i_b}'=
\left\{-1,-3,4\right\},~i=1,\ldots ,8.
\nnmb 
\eea
Such triplets are also attractive because they automatically allow 
a trilinear coupling between their component fields which
generates a superpotential mass term for each of the components.  
By explicit calculation, we find that the corresponding fermion mass 
matrix is non-singular provided all three members of a triplet develop VEV's.

  As a first application, suppose the field $Z_M$, 
with $\u1x$ charge $z_M$ but otherwise a singlet, is part of an 
anomaly-free $\u1x$ model that does not allow a superpotential 
contribution to its mass.  To generate an effective mass term 
involving $Z_M$, it is sufficient 
to introduce a set of $X_a$ and $Y_b$ fields having charges 
$\{z_M,5z_M,-6z_M\}$ and $\{-2z_M,-3z_M,5z_M\}$.  
If both $Z_M$ and some of the additional singlets get VEV's, we find 
that the full mass matrix for these fields is non-singular.  
For a second example, suppose that we wish to induce a mixing
between the fields $P$ and $R$ whose quantum numbers are such that
the product $P\,R$ is a $\gsm$ singlet but has a non-zero 
$\u1x$ charge $Q_X$.  By introducing $X_a$ and $Y_b$ singlets with
charges $\{-Q_X,-5Q_X,6Q_X\}$ and $\{2Q_X,3Q_X,-5Q_X\}$, we obtain
the desired mixing provided the $X_1$ field with charge 
$-Q_X$ condenses in the vacuum.

\section{Basic $E_6$ Facts\label{app:e6}}
  Additional $U(1)$ gauge groups may arise from the spontaneous
breaking of $E_6$.  This can happen through the sequence~\cite{Hewett:1988xc}
\beq
E_6 ~\to~ SO(10)\times U(1)_{\psi} ~\to~ SU(5)\times U(1)_{\chi}
\times U(1)_{\psi}.
\eeq
Another attractive feature of $E_6$ models is that all the matter fields
of the MSSM can be embedded in the $\mathbf{27}$ of this group.
The full particle content of the $\mathbf{27}$, as well as the 
charges under $\gsm$ and $U(1)_{\psi,\chi}$ subgroups are given
in Table~\ref{tab:1}.

\begin{table}
\caption{\label{tab:1} Particle content and subgroup charges of
the $\mathbf{27}$ representation of $E_6$~\cite{Hewett:1988xc}.}
\begin{ruledtabular}
\begin{tabular}{cccc}
$\mathbf{27}$&$\gsm$&$2\sqrt{6}\,U(1)_{\psi}$&$2\sqrt{10}\,U(1)_{\chi}$\\
\hline
$Q$&$(\mathbf{3},\mathbf{2},1/6)$&1&-1\\
$L$&$(\mathbf{1},\mathbf{2},-1/2)$&1&3\\
$U^c$&$(\mathbf{\bar{3}},\mathbf{1},-2/3)$&1&-1\\
$D^c$&$(\mathbf{\bar{3}},\mathbf{1},1/3)$&1&3\\
$E^c$&$(\mathbf{1},\mathbf{1},1)$&1&-1\\
$N^c$&$(\mathbf{1},{1},0)$&1&-5\\
\hline
$H$&$(\mathbf{1},\mathbf{2},-1/2)$&-2&-2\\
$P^c$&$(\mathbf{\bar{3}},\mathbf{1},1/3)$&-2&-2\\
$H^c$&$(\mathbf{1},\mathbf{2},1/2)$&-2&2\\
$P$&$(\mathbf{{3}},\mathbf{1},-1/3)$&-2&2\\
$S$&$(\mathbf{1},\mathbf{1},0)$&4&0
\end{tabular}
\end{ruledtabular}
\end{table}

\comment{
\beq
\begin{array}{|c|c|c|c|}
\hline
\mathbf{27}&\gsm&2\sqrt{6}\,U(1)_{\psi}&2\sqrt{10}\,U(1)_{\chi}\\
\hline
Q&(\mathbf{3},\mathbf{2},1/6)&1&-1\\
L&(\mathbf{1},\mathbf{2},-1/2)&1&3\\
U^c&(\mathbf{\bar{3}},\mathbf{1},-2/3)&1&-1\\
D^c&(\mathbf{\bar{3}},\mathbf{1},1/3)&1&3\\
E^c&(\mathbf{1},\mathbf{1},1)&1&-1\\
N^c&(\mathbf{1},{1},0)&1&-5\\
\hline
H&(\mathbf{1},\mathbf{2},-1/2)&-2&-2\\
H^c&(\mathbf{1},\mathbf{2},1/2)&-2&2\\
P^c&(\mathbf{\bar{3}},\mathbf{1},1/3)&-2&-2\\
P&(\mathbf{{3}},\mathbf{1},-1/3)&-2&2\\
S&(\mathbf{1},\mathbf{1},0)&4&0\\
\hline
\end{array}\nnmb
\eeq
Under $SU(5)$, the $\mathbf{27}$ decomposes as 
$\mathbf{\bar{5}}\oplus\mathbf{\overline{10}}\oplus\mathbf{\bar{5}}
\oplus\mathbf{5}\oplus\mathbf{1}\oplus\mathbf{1}$.
}

{\bf Acknowledgments:}
This work was supported by the Department of Energy.
We would like to thank S.P~Martin for useful comments
on the manuscript.


\end{document}